\documentclass[conference]{IEEEtran}
\IEEEoverridecommandlockouts
\usepackage{cite}
\usepackage{siunitx}
\usepackage{amsmath,amssymb,amsfonts}
\usepackage{algorithmic}
\usepackage{graphicx}
\usepackage{textcomp}
\usepackage{booktabs}
\usepackage{xcolor}
\usepackage{bm}
\usepackage{esvect}
\def\BibTeX{{\rm B\kern-.05em{\sc i\kern-.025em b}\kern-.08em
    T\kern-.1667em\lower.7ex\hbox{E}\kern-.125emX}}
\begin{document}

\title{Reconstruction-based Out-of-Distribution Detection for Short-Range FMCW Radar}

%

\author{\IEEEauthorblockN{Sabri Mustafa Kahya$^{\star}$ \qquad Muhammet Sami Yavuz$^{\star \dagger}$ \qquad Eckehard Steinbach$^{\star}$}

\IEEEauthorblockA{$^{\star}$Technical University of Munich,
School of Computation, Information and Technology,\vspace{-0.03cm}\\  Department of Computer Engineering, Chair of Media Technology,\vspace{-0.03cm}\\ Munich Institute of Robotics and Machine Intelligence (MIRMI) \\ $^{\dagger}$Infineon Technologies AG}

}

\maketitle

\begin{abstract}

 Out-of-distribution (OOD) detection recently has drawn attention due to its critical role in the safe deployment of modern neural network architectures in real-world applications. The OOD detectors aim to distinguish samples that lie outside the training distribution in order to avoid the overconfident predictions of machine learning models on OOD data. Existing detectors, which mainly rely on the logit, intermediate feature space, softmax score, or reconstruction loss, manage to produce promising results. However, most of these methods are developed for the image domain. In this study, we propose a novel reconstruction-based OOD detector to operate on the radar domain. Our method exploits an autoencoder (AE) and its latent representation to detect the OOD samples. We propose two scores incorporating the patch-based reconstruction loss and the energy value calculated from the latent representations of each patch. We achieve an AUROC of 90.72\% on our dataset collected by using 60 \si{\textbf{\GHz}} short-range FMCW Radar. The experiments demonstrate that, in terms of AUROC and AUPR, our method outperforms the baseline (AE) and the other state-of-the-art methods. Also, thanks to its model size of 641 kB, our detector is suitable for embedded usage.

\end{abstract}


\section{Introduction}
In recent years, modern deep learning models have set the state-of-the-art (SOTA) in various applications. However, they tend to give high confidence to out-of-distribution (OOD) samples. For instance, during the inference of a classification task, an instance can be assigned to one of the classes with an overconfident prediction even though the sample does not belong to any class that appears in the training set. Since these models make closed-world assumptions, their predictions on real-world scenarios such as medical and autonomous driving applications may cause undesirable consequences. Hence, to mitigate the overconfident predictions on OOD data, various OOD detection studies \cite{b24, b25, b26, b27, b28, b29, b30, b31} have been revealed. The same problem also exists for radar-based applications.

Radars, thanks to their insensitivity to illumination and environmental conditions like fog and rain, and their ability to preserve privacy, have attracted attention for several applications such as presence sensing \cite{b32}, human activity classification \cite{b34}, and gesture recognition \cite{b33}. However, none of the previous works consider how to act when a sample from a different distribution comes into evaluation. For instance, in \cite{b35}, the authors aim to detect abnormal human activities like fainting and falling in a toilet cabin. However, they do not consider when a non-human moving object like a robot vacuum cleaner appears in the toilet. Typically, the system needs to say, "unknown object," but it says either normal or abnormal activity, which belongs only to human targets. This study aims to create an OOD detector that works with Range Doppler Images (RDIs) of 60 \si{\GHz} L-shaped FMCW Radar. The OOD detector tries to detect moving objects other than a walking human.

An OOD detector works as follows; It first assigns a score to the test sample \textbf{x} by using a score function $S$. Then, using the validation data containing only in-distribution (ID) samples, a threshold $\tau$ is calculated. Finally, if $S(\textbf{x}) < \tau$, the sample is treated as ID, otherwise as OOD. To determine the scores, some detectors rely on softmax probabilities \cite{b1, b2}, some use the logits \cite{b7}, and some exploit intermediate feature representations \cite{b4,b5, b6 ,b31}. In this work, we propose two scores: Patch-based reconstruction (PB-REC) and latent energy scores (PB-LSE). For PB-REC, we use a patch-based autoencoder architecture. For PB-LSE, on the other hand, we use the latent representation of the autoencoder by bringing it from a 1D vector $(\mathbb{R}^D)$ to a scalar $(\mathbb{R})$. With the help of patch-based training (only with ID samples) and inference strategy, we focus on the local features of RDIs. Thus, the local features of an ID sample are better reconstructed, leading to better OOD detection during inference. Also, with the same idea, for a single RDI, we get as many latent codes as patches in inference time, which means that we get compressed but more detailed information from each patch. We combine the latent codes to get the final energy score (PB-LSE). This combination makes the assigned energy score more distinguishable for ID and OOD samples since we have more information about the data. The experiments in Section \ref{experiments} also reflect the correctness of the proposed idea. Our key contributions are as follows:
\begin{itemize}

    \item We propose a reconstruction-based OOD detector. Our detector uses two scoring functions that produce patch-based reconstruction and latent energy scores. We achieve an AUROC of 90.72\% on our dataset, consisting of several ID and OOD samples.
    
    \item Compared to the baseline and the other SOTA methods, our method has superior results. Besides, when we only use the energy scores, we get an AUROC of 90.72\%. Thus, we can differentiate between ID and OOD samples without needing the decoder, using only the encoder part of the network. The size of the encoder part is only 641 kB, making our detector very suitable for embedded usage.
    
\end{itemize}

\section{Related Work}
The existing studies on OOD detection are mainly shaped by the image domain, especially for classification tasks. In this regard, as a baseline method, Hendrycks \& Gimpel published their work \cite{b1}, which is based on maximum softmax probability (MSP). Their main claim is that in the inference of deep neural networks, higher confidence values are given to ID samples compared to OOD samples. Therefore, with simple thresholding, the ID and OOD samples can be differentiated. The ODIN approach introduced in \cite{b2} improves \cite{b1} by introducing temperature scaling and input perturbations to increase the softmax scores of ID samples. This method is extended by introducing a model ensembling approach \cite{b3}. Mahalanobis distance based detector (MAHA) \cite{b4} aims to detect OOD samples using intermediate feature representations of DNNs. It creates a class conditional Gaussian distribution for each ID class and assigns an OOD score to the test sample using the Mahalanobis distance of the sample to each of the class conditional Gaussian distributions. It also uses layer ensembling and input perturbation like ODIN \cite{b2}. Similarly, \cite{b5, b6} utilize intermediate representations to detect OOD samples. Liu et al. \cite{b7} created an energy-based OOD detection method by applying the $LogSumExp(LSE)$ function on the logit layer and stated that ID samples have lower energy scores than OOD samples. The methods above work with any pre-trained model, but some of them require OOD samples to tune the hyper-parameters.

In the Outlier Exposure (OE) technique by Hendrycks et al. \cite{b8}, a limited number of OOD samples are used for either fine-tuning or training from scratch with a new loss, which pushes the softmax probabilities of the OODs to the uniform distribution. In the inference step, they only use regular cross-entropy loss and propose a score-based OOD detector. Similar studies \cite{b9, b10} exploit OE while training their deep architectures with novel loss functions. In \cite{b11}, a GAN architecture is used with a special loss function to generate artificial OOD samples to be used for OE. In the literature, many other studies \cite{b12, b13} utilize generative models like GANs and Normalizing Flows to detect OODs.

Some studies \cite{b2, b14} emphasize the importance of gradient information for the OOD detection task. For example, in GradNorm \cite{b14}, the magnitudes of the gradients are utilized to distinguish between IDs and OODs. The gradients are calculated by backpropagating the KL divergence between the softmax output and uniform distribution. In this logic, the ID samples have higher magnitudes than OODs.

There are also some reconstruction-based OOD detection methods. For example, \cite{b15} uses AE, and as an OOD score, it combines the reconstruction loss with a Mahalanobis distance in the latent code. Some other studies \cite{b16 ,b17} benefit from Variational Autoencoder (VAE) and its latent representation to detect OODs.

Most of the works in this area are based on convolutional neural network (CNN) architectures. However, in recent years, some works \cite{b18, b19, b20} based on pre-trained transformers with attention mechanisms like VIT \cite{b21} and BERT \cite{b22} are revealed as OOD detectors, and they achieved SOTA results on some benchmark datasets.

In the radar domain, there are few OOD detection studies utilizing DNNs. One study \cite{b23} compares a few conventional and deep learning methods for low-resolution radar micro-Doppler signatures using a synthetically generated dataset. In \cite{b36}, a hand gesture recognition system is proposed that leverages FMCW radar technology and includes the ability to detect OOD input data. \cite{b37} proposes a meta-reinforcement learning approach for robust radar tracking with OOD detection support.

\section{System Design}
\subsection{Radar Configuration}
Our sensor is based on Infineon's BGT60TR13C chipset with a 60 \si{\GHz} L-Shaped FMCW radar. It consists of one transmit (Tx) and three receiver (Rx) antennas. The radar configuration is listed in Table \ref{tab:radar_conf}. The Tx antenna transmits $N_c$ chirp signals, and the Rx antennas receive the reflected signals. The transmitted and reflected signals are mixed, and the mixture of the signals produces the intermediate frequency (IF). The final raw Analogue-to-Digital Converter (ADC) data is acquired by low-pass filtering and digitizing the IF signal. Each chirp has $N_s$ samples, so with some rearranging, the dimensions of a frame are shaped as $N_{Rx} \times N_c \times N_s$, and the frame becomes ready for further digital signal processing.

\subsection{Pre-processing}
In our network, we use RDIs as inputs; therefore, we apply the following pre-processing steps to our raw data:
\begin{itemize}
    \item \textbf{Range FFT}: We apply range FFT with Chebyshev window at 100 \si{\dB} on fast time with mean removal to get one channel range data out of the three channels from three Rx antennas.
    \item \textbf{MTI}: Following, we apply simple frame-wise moving target identification (MTI) on range data to remove any static object in the field of view of the radar.
    \item \textbf{Doppler FFT}: We then perform doppler FFT with Chebyshev window at 100 \si{\dB} on slow time for each sample of a chirp and have our final RDI with the dimension of $64 \times 64$.

\end{itemize}

\begin{table}[h]
    \caption{\small FMCW Radar Configuration Parameters }
    \centering
    \begin{tabular}{@ {\extracolsep{10pt}} ccc}
    \toprule

    \centering
    Configuration name & Symbol & Value \\
    \midrule
    Number of Transmit Antennas & $N_{Tx}$  & 1  \\
    Number of Receive Antennas & $N_{Rx}$  & 3  \\
    Sampling frequency & $f_s$ & 2 \si{\MHz} \\
    Number of chirps per frame & $N_c$ & 64  \\
    Number of samples per chirp & $N_s$  & 128 \\
    Frame Period & $T_f$ & 50 \si{\ms}  \\
    
    Chirp to chirp time & $T_c$ & 391.55 \si{\us} \\
    Ramp start frequency & $f_{min}$ & 60.1 \si{\GHz}\\ Ramp stop frequency & $f_{max}$ & 61.1 \si{\GHz}\\
    Bandwidth & $B$ & 1 \si{\GHz}\\ 
    \bottomrule
    \end{tabular}

    \label{tab:radar_conf}
\end{table}

\begin{figure}[htbp]
\centerline{\includegraphics[width=\linewidth]{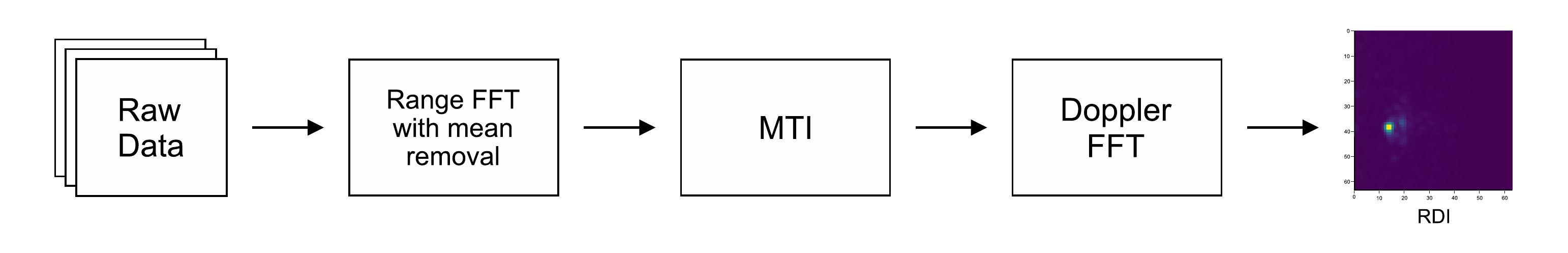}}
\caption{The diagram of the pre-processing steps to obtain RDIs}
\label{fig:preprocess}
\end{figure}

\section{Problem Statement and Method}
In this work, we aim to have a system that can identify and appropriately respond to moving targets (OOD), except for a single walking human (ID) using the RDIs.
Normally, this is a binary (0-1) classification task. The ID samples should be classified as one, while the OODs need to be labeled as zero. However, the OOD detection task differs from a typical binary classification application since there may be infinitely many OOD objects, which need to be correctly detected. That is why we test our pipeline with a large number of distinct OOD objects, which may appear with a high chance in indoor environments like offices and homes. We are also confident that our pipeline would perform well when tested with different OODs we do not use in our study.

\subsection{Architecture and Training}
For our reconstruction-based OOD detection pipeline, we use a patch-based AE architecture. The AE aims first to compress the input data to a latent space with smaller dimensions than the input sample, and from the latent space, it tries to reconstruct the input without losing information.

Our patch-based AE architecture consists of encoder and decoder parts. The encoder includes three 2D convolution layers with ReLu activation and a pooling layer after each convolution layer. Each convolution has a kernel size of $3 \times 3$. The first has 16 filters while the others have 32 and 64, respectively. With a flattening and a dense layer, the encoder brings the input to its latent space (128). The decoder, on the other hand, first takes the latent code and sequentially feeds it to dense and reshaping layers. Then it applies four transpose 2D convolutions with an up-sampling layer after the first three convolution layers. The first three convolution layers have ReLu activation, and a kernel size of $3 \times 3$. The first has 64 filters, while the others have 32 and 16. To reconstruct the input with the same dimensionality, the final convolution layer consists of a $3 \times 3$ kernel size with one filter, and sigmoid activation.  

For the training, we only use ID samples. We divide each RDI (64,64,1) into four equally sized (32,32,1) patches. These patches are individually fed into our network while preserving their original spatial arrangement within one compact, unified RDI representation. With this technique, the system focuses on the local features instead of learning the entire RDI once. Together with Adam optimizer, we use binary cross-entropy as the loss function and calculate it between the input RDI and the concatenated reconstructions.

\subsection{OOD Detection}
\begin{figure}[htbp]
\centerline{\includegraphics[width=\linewidth]{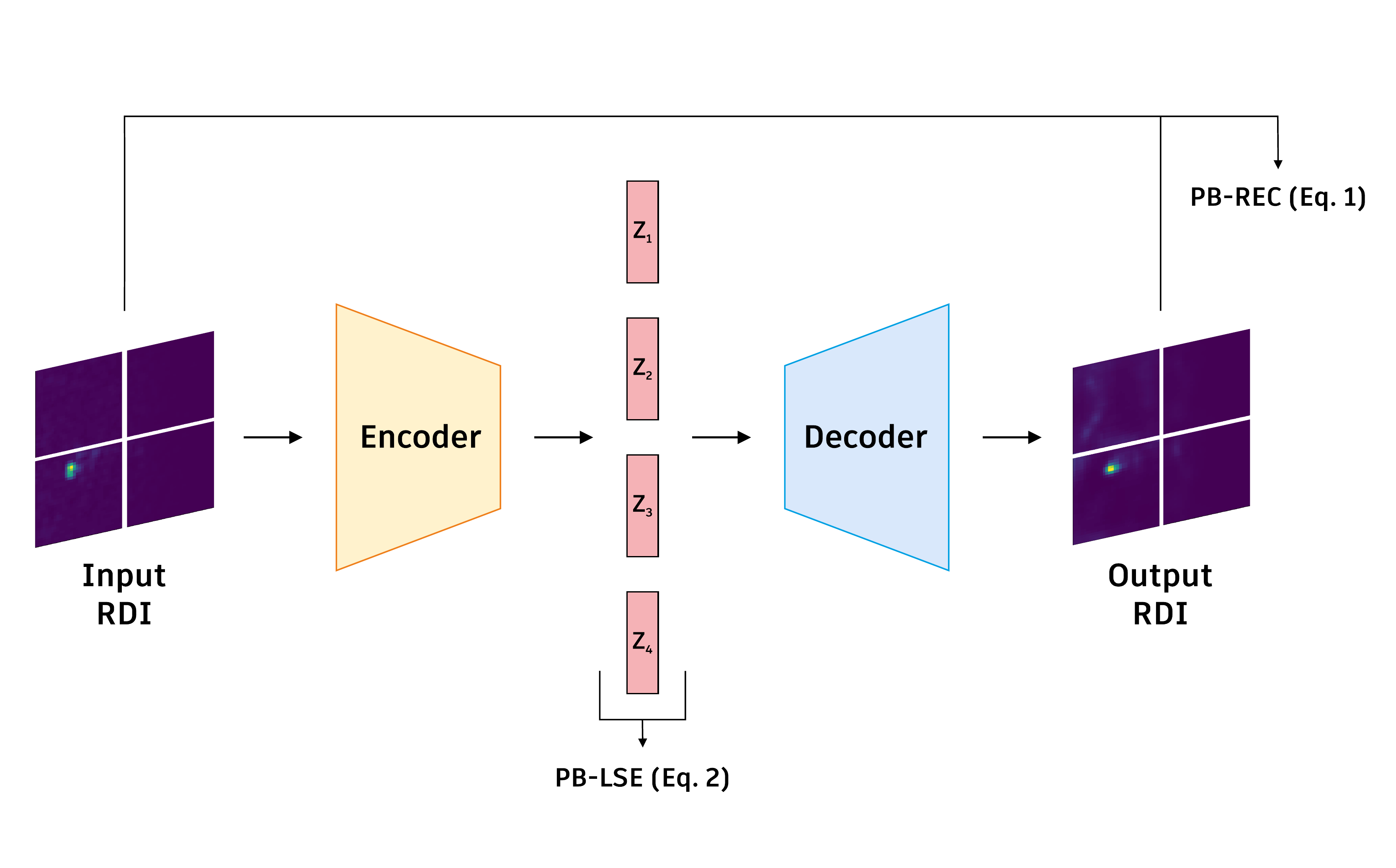}}
\caption{The overall pipeline of the proposed method.}
\label{fig:pipline}
\end{figure}

To detect the OOD data, we propose two scores consisting of reconstruction and latent energy scores. During inference, each instance in the RDI sequence is again divided into four patches, and each patch is fed to the pre-trained DNN model. The entire model gives the reconstruction results from each patch. We combine the reconstructed patches based on their position and calculate the mean of the reconstruction mean squared error (MSE). This score represents the reconstruction score (\ref{eq:1}). The encoder part of the model gives four latent codes coming from each patch. We first sum them and apply the LogSumExp ($LSE$) function to the summed latent code (1D vector) and calculate a latent energy score out of it (\ref{eq:2}). Then we simply use the scores to assess the performance of our method on different metrics. The overall pipeline is presented in Figure \ref{fig:pipline}.

\vspace{-0.40cm}
\begin{equation}
   \small {S_r(\textbf{X}) = \frac{1}{q}\sum_{z=1}^q\frac{1}{hw} \sum_{i=1}^h \sum_{j=1}^w (D(E(\textbf{P}_z))_{ij}-(\textbf{P}_z)_{ij})^2 }
    \label{eq:1}
\end{equation}

\begin{equation}
    \small {S_e(\textbf{X}) =  \log\left[\sum_{j=1}^k \exp(\sum_{t=1}^q(\textbf{z}_t(j))) \right]}
    \label{eq:2}
\end{equation}
where $\textbf{X} \in \mathbb{R}^{64 \times 64}$ is the input RDI and consists of four patches $\{\textbf{P}_1,\textbf{P}_2,\textbf{P}_3,\textbf{P}_4\}$. $E$ and $D$ are encoder and decoder respectively. $\textbf{z}_t(j)$ is the j-th element of $\textbf{z} \in \mathbb{R}^{k}$ for patch $\textbf{P}_t$. $h$ and $w$ are the dimensions of a patch (32,32), $q$ is the number of patches (4), and $k$ is the length of the latent vector (128). 

Energy-based models (EBM) are used in the literature for OOD tasks. They mainly depend on a function that maps $N_d$ dimensional data vectors to a scalar value called energy. \cite{b7} utilizes this idea by applying LSE on the logit layer, which involves class label information of a K-class classification task. We are inspired by \cite{b7} and apply LSE on the latent code that has all the information about the input in a compressed way. The main difference is that they have classification-based architectures and get the energy scores just before the softmax layer, which includes information from all classes, including the actual class the data belongs to. On the other hand, we use reconstruction-based architecture and have our energy scores from the latent representation of an AE. Meaning our energy scores come from the real data and only the actual class and are expected to be more informative. We expect lower energy scores for ID samples than for OOD samples, as well as reconstruction scores due to the fact that we train our network only with ID samples.

\section{Experiments}
\label{experiments}
\subsection{Dataset}
The RF data is collected using Infineon's BGT60TR13C 60 \si{\GHz} FMCW radar sensor with four individuals in 10 distinct indoor places, including office and house rooms. The radar is placed at two meters height and tilted 15 degrees down from the front side. Our dataset consists of ID and OOD samples. ID samples comprise a human walking activity. For this scenario, the person in the radar field of view freely walks without supervision with a changing distance from 1 to 5m. The free walks contain any movement, such as forward-backward walks, from left to right and right to left, and circular and diagonal walks. On the other hand, OOD samples consist of various moving objects seen in indoor environments, including a table fan, a remote-controlled (RC) toy car, swinging plants, swinging blinds, swinging laundry, a swinging paper bag, a vacuum cleaner, and a robot vacuum cleaner. Each activity is recorded for around 1 to 2 minutes. After the pre-processing, we obtain a set of RDI data from the recordings of 10 separate rooms and use six rooms for the training and the remaining four for the test phase. For training, we use 20000 ID frames from six different rooms and four different individuals. These frames consist of a balanced number of individuals and rooms. For inference, we exploit many ID (10000) and OOD (12000) samples from four different rooms with a balanced number of individuals, other moving objects, and rooms. For similar radar-based applications, DL models tend to learn the room itself, so in some cases, an application may perform well in one room but poorly in different rooms. This means the system may overfit to the room. In our experiments, none of the rooms utilized during training appear in testing, and this reflects the robustness of our pipeline. The dataset details are provided in Table \ref{tab:data}.

\begin{table}[h]
    \caption{\small Number of ID and OOD RDI data samples for each activity used in the training and test.  }
    \centering
    \begin{tabular}{@ {\extracolsep{4pt}} c|c|c}
    \toprule
    \centering
    ID Activity  & \# of Training Samples & \# of Test Samples \\
    \midrule
    Walking & 20000 & 10000\\

    \midrule
    \midrule
    OOD Activity  \\
    \midrule
     Table Fan & -&2000 \\
    RC Toy Car & -&2000 \\
    Swinging Laundry & -&2000\\
    Robot Vacuum Cleaner &-  & 1500\\
    Vacuum Cleaner & -&1500 \\
    Swinging Blinds & -&1000\\
    Swinging Plants & -&1000\\
    Swinging Paper Bag & -&1000\\    
  
    \midrule
    Total OOD & - & 12000 \\
    \midrule
    Total & 20000& 22000 \\
    
    \bottomrule
    \end{tabular}

    \label{tab:data}
\end{table}

\vspace{-0.1cm}
\subsection{Evaluation Metrics and Experimental Results}
In OOD tasks, accuracy is not a reliable metric to be used, so researchers commonly evaluate their OOD detectors using the following metrics:
\begin{itemize}
    \item \textbf{AUROC} is the area under the receiver operating characteristic (ROC) curve. 
    \item \textbf{AUPR\_IN} is defined as  the area under the precision-recall curve when ID samples are assumed to be positives.  
      \item \textbf{AUPR\_OUT} is the area under the precision-recall curve when OOD samples are assumed to be positives.
  
\end{itemize}

\begin{table}[h]
    \caption{\small \textbf{Main Results.} Comparison with the baseline (AE) and other SOTA methods. All values are percentages.}
    \centering
    
    \begin{tabular}{@ {\extracolsep{10pt}} cccc}
    \toprule
    
    \centering
    Method & AUROC & AUPR\_IN & AUPR\_OUT\\
    \midrule
       Baseline (AE) & 75.44 &  78.07 & 76.80  
       \\ \midrule
       MSP (LeNet5) \cite{b1} & 88.66 & 81.03 & 93.26\\
       Energy (LeNet5)  \cite{b7} & 88.85 & 82.49&93.31 \\
       ODIN (LeNet5)  \cite{b2} & 88.86 & 82.50 & 93.32 \\ 
       OE (LeNet5)  \cite{b8} & 50.34 & 32.31 & 67.80 \\ \midrule
       MSP (ResNet-34) \cite{b1} & 50.07 & 29.00 & 71.17 \\
       Energy (ResNet-34)  \cite{b7} & 68.09 & 64.90&74.29 \\
       ODIN (ResNet-34)  \cite{b2} & 68.68 & 65.35 & 74.67  \\ 
        OE (ResNet-34)  \cite{b8} & 17.38 & 20.61 & 49.69 \\ \midrule
       MSP (ResNet-50) \cite{b1} & 50.13 & 28.96 & 70.78 \\
       Energy (ResNet-50)  \cite{b7} & 89.82 & 79.48&94.36 \\
       ODIN (ResNet-50)  \cite{b2} & 89.86 & 79.51 &\textbf{ 94.40 } \\ 
       OE (ResNet-50)  \cite{b8} & 7.99 & 18.89 & 47.21  \\ \midrule
        PB-REC (Ours)  & 88.84 &  83.61 & 92.16 \\
        \textbf{PB-LSE} (Ours) & \textbf{90.72} &  \textbf{87.67} & 92.81 \\ 
    
    \bottomrule
    \end{tabular}
    \label{tab:main_results}
\end{table}

Table \ref{tab:main_results} demonstrates the superiority of the proposed method in terms of AUROC and AUPR\_IN over the baseline and other SOTA methods. The baseline method (AE) uses the same architecture as our proposed methods, and instead of using patches, it directly uses the RDIs to train the AE. In the OOD detection phase, the baseline method only uses the reconstruction scores calculated from the mean of the reconstruction MSE in between input and reconstructed RDIs. In MSP, Energy, ODIN, and OE methods, LeNet5 \cite{lecun1998gradient}, ResNet-34, and ResNet-50 \cite{resnet34} backbones are used. The architectures are trained in a one-class classification manner by only using ID samples, and the trained models are used to apply the SOTA methods. 

\section{Conclusion}
OOD detection has a crucial place in academia and industry since being able to detect unknown samples has great importance, especially for safety-critical applications like medical and autonomous driving. In the literature, there are several OOD detection methods, and they produce impressive results. However, in the radar domain, this field is not studied broadly. Therefore, in this study, we addressed the OOD detection problem on top of a short-range FMCW radar. We propose two score values by using a reconstruction score from an autoencoder and a latent energy score from the latent representation of the same autoencoder to detect the OOD samples, which are all the moving objects in the field of view of the radar other than a walking person. With our approach, we reach an AUROC of 90.72\%, which outperforms the baseline as well as other compared SOTA methods. Our experiments also demonstrate the effectiveness of patch-based strategy. Thanks to it, we get better reconstructions and more informative latent codes, resulting in more distinguishable reconstruction and energy scores for ID and OOD samples. Additionally, with its small size of 641 kB, our detector is appropriate for embedded devices. 

\bibliographystyle{ieeetr} 
\end{document}